\documentstyle[12pt]{article}
\topmargin - 0.5in
\oddsidemargin - 5mm
\begin{document}
\def\p{\phi}
\def\P{\Phi}
\def\a{\alpha}
\def\e{\varepsilon}
\def\be{\begin{equation}}
\def\ee{\end{equation}}
\def\l{\label}
\def\0{\setcounter{equation}{0}}
\def\h{\hat}
\def\b{\beta}
\def\S{\Sigma}
\def\C{\cite}
\def\r{\ref}
\def\ba{\begin{eqnarray}}
\def\ea{\end{eqnarray}}
\def\n{\nonumber}
\def\X{\Xi}
\def\x{\xi}
\def\la{\lambda}
\def\d{\delta}
\def\D{\Delta}
\def\s{\sigma}
\def\f{\frac}
\def\Ga{\Gamma}
\def\ga{\gamma}
\def\th{\theta}
\def\Th{\Theta}
\def\pa{\partial}
\def\o{\omega}
\def\O{\Omega}
\def\vae{\varepsilon}
\def\rar{\rightarrow}
\def\vp{\varphi}
\def\R{\rho}
\def\D{\Delta}

\vskip 3cm
\begin{titlepage}
\begin{center}
{\Large \bf Example of quantum systems reduction}
\vskip 1cm
J.Manjavidze \\Institute of Physics, Tbilisi,
Rep. of Georgia \\
\end{center}
\vskip 1.5cm

\begin{abstract}\footnotesize

To solve the quantum-mechanical problem the procedure of mapping onto
linear space $W$ of generators of the (sub)group violated by given
classical trajectory is formulated. The formalism is illustrated by
the plane H-atom model. The problem is solved noting conservation of
the Runge-Lentz vector $n$ and reducing the 4-dimensional incident
phase space $T$ to the 3-dimensional linear subspace $W=T^* V\times
R^1$, where $T^* V$ is the (angular momentum ($l$) - angle ($\vp$))
phase space and $R^1 =n$. It is shown explicitly that (i) the motion
in $R^1$ is pure classical as the consequence of the reduction, (ii)
motion in the $\vp$ direction is classical since the Kepler orbits
are closed independently from initial conditions and (iii) motion in
the $l$ direction is classical since all corresponding quantum
corrections are defined on the bifurcation line ($l=\infty$) of the
problem.  In our terms the H-atom problem is exactly quasiclassical
and is completely integrable by this reasons.

\end{abstract}
\end{titlepage}
\section{Introduction}\0

The mapping
\be
J:T \rar W,
\l{11}\ee
where $T$ is the $2N$-dimensional phase space and $W$ is a linear
space solves the mechanical problem iff
\be
J=\otimes^N_1 J_i,
\l{12}\ee
where $J_i$ are the first integrals in involution, e.g. \C{arn} (the
formalism of reduction (\r{11}) in classical mechanics is described
also in \C{mars}). The aim of this article is to adopt this procedure
for quantum systems.

The mapping (\r{11}) introduces integral $manifold$
$J_{\o}=J^{-1}(\o)$ in such a way that the $classical$ phase space
flaw with given $v \in J_{\o}$ belongs to $J_{\o}$ $completely$. We
wish quantize the $J_{\o}$  manifold instead of flow in $T$ noting
that the quantum trajectory also should belong to $J_{\o}$
completely. This important conclusion was demonstrated in \C{yaph} by
the canonical transformation of the path-integral measure. New
perturbation theory is extremely simple since $W$ is the linear
space.

The `direct' mapping (\r{11}) used in \C{yaph} assumes that $J$ is
known.  But it seems inconvenient having in mind the general problem
of nonlinear waves quantization, when the number of degrees of
freedom $N=\infty$, or if the transformation is not canonical. We
will consider by this reason the `inverse' approach assuming that the
classical flow is known. Then, since the flow belongs to $J_{\o}$
completely \C{yaph}, we would be able to find the quantum motion in
$W$. It is the main technical result illustrated in this paper.

The manifold $J_{\o}$ is invariant relatively to some subgroup
$G_{\o}$ \C{smale} in accordance to topological class of classical
flaw. This introduces the $J_{\o}$ classification and summation over
all (homotopy) classes should be performed. Note, the classes are
separated by the boundary bifurcation lines in $W$ \C{smale}. If the
quantum perturbations switched on adiabatically then the homotopy
group should stay unbroken. It is the ordinary statement for quantum
mechanics. (But, generally speaking, this is not true for field
theories.)

We will calculate the bound state energies in the Coulomb potential.
This popular problem was considered by many authors, using various
methods, see, e.g., \C{pop}. The path-integral solution of this
problem was offered in \C{dur}. We will restrict ourselves by the
plane problem. Corresponding phase space $T=(p,l,r,\vp)$ is
4-dimensional.

The classical flaw of this problem can be parametrized by the angular
momentum $l$, corresponding angle $\vp$ and by the normalized on
total Hamiltonian Runge-Lentz vector length $n$. So, we will consider
the mapping ($p$ is the radial momentum in the cylindrical
coordinates):
\be
J_{l,n}:(p,l,r,\vp)\rar(l,n,\vp)
\l{15}\ee
to construct the perturbation theory in the $W=(l,n,\vp)$ space. I.e.
$W$ is not considered as the cotangent foliation on $T$.

The mapping (\r{15}) assumes additional reduction of the
four-dimensional incident phase space up to three-dimensional linear
subspace\footnote{$W$ would not have the symplectic structure.}. Just
this reduction phenomena leads to corresponding stability of $n$
concerning quantum perturbations and will allow to solve our H-atom
problem completely\footnote{ In other words, we would demonstrate
that the hidden Bargman-Fock \C{pop} $O(4)$ symmetry is stay unbroken
concerning quantum perturbations.}.

In Sec.2 we will show how the mapping (\r{15}) can be performed for
path-integral differential measure. In Sec.3 the consequence of
reduction will be derived and in Sec.4 the perturbation theory in the
$W$ space will be analyzed. The calculations are based on the
formalism offered in \C{yaph}.

\section{Mapping}\setcounter{equation}{0}

We will calculate the integral \C{yaph}:
\be
\R(E)=\int^{\infty}_0 dTe^{-i\h{K}(j,e)}\int DM(p,l,r,\vp)
e^{-iV(r,e)},
\l{21}\ee
where $\R(E)$ is the $probability$ to find a particle with energy
$E$, i.e.  we should find \C{manj} that normalized on the zero-modes
volume
\be
\R(E)=\pi\sum_n \d (E-E_n),
\l{22}\ee
where $E_n$ are the bound states energies. For $H$-atom problem
$E_n\leq 0$.  This condition will define considered homotopy class.

Expansion over operator
\be
\h{K}(j,e)=\f{1}{2}\int^T_0 dt (\h{j}_r\h{e}_r +
\h{j}_{\vp}\h{e}_{\vp}),~~~
\h{X}(t)\equiv \d/\d X(t),
\l{23}\ee
generates the perturbation series. It will be seen that in our case
we may omit the question of perturbation theory convergence.

The differential measure
\ba
DM(p,l,r,\vp)=\d (E-H_0)\prod_t dr(t) dp(t) dl(t) d\vp (t)
\d (\dot{r}-\f{\pa H_j}{\pa p})\times
\n \\ \times
\d (\dot{p}+\f{\pa H_j}{\pa r})
\d (\dot{\vp}-\f{\pa H_j}{\pa l})
\d (\dot{l}+\f{\pa H_j}{\pa \vp}),
\l{24}\ea
with total Hamiltonian ($H_0=H_j|_{j=0}$)
\be
H_j=\f{1}{2}p^2 -\f{l^2}{2r^2}-\f{1}{r}-j_r r -j_{\vp}\vp
\l{25}\ee
allows perform arbitrary transformations because of its
$\d$-likeness. The functional
\be
V(r,e)=-s_0(r)+\int^T_0 dt[\f{1}{((r+e_r)^2+r^2e_{\vp}^2)^{1/2}}-
\f{1}{((r-e_r)^2+r^2e_{\vp}^2)^{1/2}}+2\f{e_r}{r}]
\l{26}\ee
describes the interaction between various quantum modes and $s_0 (r)$
defines the nonintegrable phase factor \C{manj}. The quantization of
this factor determines the bound state energy (see below).  Such
factor will appear if the phase of amplitude can not be fixed (as,
for instance, in the Aharonov-Bohm case).  Note that the Hamiltonian
(\r{25}) contains the energy of radial $j_r r$ and angular
$j_{\vp}\vp$ excitation independently.

Let
\be
\D=\int \prod_t d^2\x d^2\eta \d (r-r_c(\x,\eta))\d (p-p_c(\x,\eta))
\d (l-l_c(\x,\eta))\d (\vp-\vp_c(\x,\eta))
\l{27}\ee
be the functional of known functions $(r_c,p_c,\vp_c,l_c)(\x,\eta)$.
It is assumed that there are such functions $(\x,\eta)(t)$ at given
$(r,p,\vp,l)(t)$ that the functional determinant
\ba
\D_c=\int \prod_t d^2\bar{\x} d^2\bar{\eta}
\d(\f{\pa r_c}{\pa\x}\cdot\bar{\x}+\f{\pa r_c}{\pa\eta}
\cdot\bar{\eta})\d(\f{\pa p_c}{\pa\x}\cdot\bar{\x}+
\f{\pa p_c}{\pa\eta}\cdot\bar{\eta})\times
\n \\
\times
\d(\f{\pa \vp_c}{\pa\x}\cdot\bar{\x}+\f{\pa \vp_c}{\pa\eta}
\cdot\bar{\eta})\d(\f{\pa l_c}{\pa\x}\cdot\bar{\x}+
\f{\pa l_c}{\pa\eta}\cdot\bar{\eta})\neq 0.
\l{28}\ea
Note that this is the condition for $(r_c,p_c,\vp_c,l_c)(\x,\eta)$
only since one can choose $(r,p,\vp,l)(t)$ in eq.(\r{27}) in an
arbitrary useful way.

To perform the mapping we should insert
$$
1=\D/\D_c
$$
into (\r{21}) and integrate over $r(t)$, $p(t)$, $\vp(t)$ and $l(t)$.
In result we find the measure:
\ba
DM(\x, \eta)=\f{1}{\D_c}\d (E-H_0)\prod_t d^2\x d^2\eta
\d (\dot{r_c}-\f{\pa H_j}{\pa p_c})\times
\n \\ \times
\d (\dot{p_c}+\f{\pa H_j}{\pa r_c})
\d (\dot{\vp_c}-\f{\pa H_j}{\pa l_c})
\d (\dot{l_c}+\f{\pa H_j}{\pa \vp_c}),
\l{29}\ea
Note that the functions $(r_c,p_c,\vp_c,l_c)(\x,\eta)$ was not
specified.

A simple algebra gives:
\ba
DM(\x, \eta)=\f{\d(E-H_0)}{\D_c}\prod_td^2\x d^2\eta
\int \prod_t d^2\bar{\x}d^2\bar{\eta}
\n \\ \times
\d^2(\bar{\x}-(\dot{\x}-\f{\pa h_j}{\pa\eta}))
\d^2(\bar{\eta}-(\dot{\eta}+\f{\pa h_j}{\pa \x}))
\n \\ \times
\d(\f{\pa r_c}{\pa\x}\cdot\bar{\x}+\f{\pa r_c}{\pa\eta}
\cdot\bar{\eta} +
\{r_c,h_j\}-\f{\pa H_j}{\pa p_c})
\n\\\times
\d(\f{\pa p_c}{\pa\x}\cdot\bar{\x}+\f{\pa p_c}{\pa\eta}
\cdot\bar{\eta} +
\{p_c,h_j\}+\f{\pa H_j}{\pa r_c})
\n \\ \times
\d(\f{\pa \vp_c}{\pa\x}\cdot\bar{\x}+\f{\pa \vp_c}{\pa\eta}
\cdot\bar{\eta}+
\{\vp_c,h_j\}-\f{\pa H_j}{\pa l_c})
\n \\\times
\d(\f{\pa l_c}{\pa\x}\cdot\bar{\x}+\f{\pa l_c}{\pa\eta}
\cdot\bar{\eta}+
\{l_c,h_j\}+\f{\pa H_j}{\pa \vp_c}).
\l{210}\ea
The Poisson notation:
$$
\{X,h_j\}=\f{\pa X}{\pa \x}\f{\pa h_j}{\pa \eta}-
\f{\pa X}{\pa \eta}\f{\pa h_j}{\pa \x}
$$
was introduced in (\r{210}).

We will define the `auxiliary' quantity $h_j$ by following
equalities:
\ba
\{r_c,h_j\}-\f{\pa H_j}{\pa p_c}=0,~
\{p_c,h_j\}+\f{\pa H_j}{\pa r_c}=0,
\n \\
\{\vp_c,h_j\}-\f{\pa H_j}{\pa l_c}=0,~
\{l_c,h_j\}+\f{\pa H_j}{\pa \vp_c}=0.
\l{211}\ea
Then the functional determinant $\D_c$ is canceled and
\be
DM(\x, \eta)=\d(E-H_0)\prod_td^2\x d^2\eta
\d^2(\dot{\x}-\f{\pa h_j}{\pa\eta})
\d^2(\dot{\eta}+\f{\pa h_j}{\pa \x}),
\l{212}\ee
It is the desired result of transformation of the measure for given
generating functions $(r_c,p_c,\vp_c,l_c)(\x,\eta)$. In this case the
`Hamiltonian' $h_j (\x,\eta)$ is defined by four equations (\r{211}).

But there is another possibility. Let us assume that
\be
h_j (\x, \eta)=H_j (r_c, p_c, \vp_c, l_c)
\l{213}\ee
and the functions $(r_c,p_c,\vp_c,l_c)(\x,\eta)$ are unknown. Then
eqs.(\r{211}) are the equations for this functions. It is not hard to
see that the eqs.(\r{211}) simultaneously with equations fixed by
$\d$-functions in (\r{212}) are equivalent of incident equations if
the equality (\r{213}) is hold. So, incident dynamical problem was
divided on two parts. First one defines the trajectory in the $W$
space through eqs.(\r{211}). Second one defines the dynamics, i.e.
the time dependence, through the equations fixed by $\d$-functions in
the measure.

Therefore, we should consider $r_c,~ p_c,~ \vp_c,~ l_c$ as the
solutions in the $\x,~\eta$ parametrization. The desired
parametrization of classical orbits has the form (one can find it in
arbitrary textbook of classical mechanics):
\be
r_c=\f{\eta_1^2(\eta_1^2+\eta_2^2)^{1/2}}
{(\eta_1^2+\eta_2^2)^{1/2}+\eta_2\cos \x_1},~
p_c=\f{\eta_2\sin \x_1}{\eta_1(\eta_1^2+\eta_2^2)^{1/2}},~
\vp_c=\x_1,~l_c=\eta_1.
\l{214}\ee
At the same time,
\be
h_j=\f{1}{2(\eta_1^2+\eta_2^2)^{1/2}} -j_r r_c -j_\vp \x_1
\equiv h (\eta)-j_r r_c -j_\vp \x_1.
\l{215}\ee

Noting that the derivatives over $\x_2$ are equal to zero\footnote{To
have the condition (\r{28}) we should assume that $\pa r_c/\pa \x_2
\sim \e \neq 0$. We put $\e =0$ completing the transformation.} we
find that
\ba
DM(\x, \eta)=\d(E-h(T))\prod_td^2\x d^2\eta
\d(\dot{\x}_1-\o_1+j_r\f{r_c}{\pa\eta_1})
\n \\ \times
\d(\dot{\x}_2-\o_2+j_r\f{r_c}{\pa\eta_2})
\d(\dot{\eta}_1-j_r\f{\pa r_c}{\pa \x_1} -j_\vp)
\d(\dot{\eta}_2),
\l{217}\ea
where
\be
\o_i=\pa h/\pa\eta_i
\l{218}\ee
are the conserved in classical limit $j_r=j_\vp =0$ `velocities' in
the $W$ space.

\section{Reduction}\0

We see from (\r{217}) that the length of Runge-Lentz vector is not
perturbated by the quantum forces $j_r$ and $j_{\vp}$. To investigate
the consequence of this fact it is useful to project this forces on
the axis of $W$ space. This means splitting of $j_r,~j_{\vp}$ on
$j_\x,~j_\eta$.  The equality
$$
\prod_t\d(\dot{\x}_1-\o_1+j_r\f{r_c}{\pa\eta_1})=
e^{\f{1}{2i}\int^T_0 dt \h{j}_{\x_1}\h{e}_{\x_1}}
e^{2i\int^T_0 dt j_r e_{\x_1}\pa r_c/\pa \eta_1}
\prod_t\d(\dot{\x}_1-\o_1+j_{\x_1})
$$
becomes evident if the Fourier representation of $\d$-function is
used (see also \C{yaph}). The same transformation of arguments of
other $\d$-functions in (\r{217}) can be applied. Then, noting that
the last $\d$-function in (\r{217}) is source-free, we find the same
representation as (\r{21}) with
\be
\h{K}(j,e)=\int^T_0 dt (\h{j}_{\x_1}\h{e}_{\x_1}+
\h{j}_{\x_2}\h{e}_{\x_2}+
\h{j}_{\eta_1}\h{e}_{\eta_1}),
\l{31}\ee
where the operators $\h{j}$ are defined by the equality:
\be
\h{j}_X (t)=\int^T_0 dt' \theta(t- t')\h{X}(t')
\l{44}\ee
and $\theta(t- t')$ is the Green function of our perturbation theory
\C{yaph}.

We should change also
\be
e_r\rar e_c=e_{\eta_1}\f{\pa r_c}{\pa \x_1}-
e_{\x_1}\f{\pa r_c}{\pa \eta_1}-
e_{\x_2}\f{\pa r_c}{\pa \eta_2},~~e_\vp\rar e_{\x_1}
\l{32}\ee
in the eq.(\r{26}). The differential measure takes the simplest form:
\ba
DM(\x, \eta)=\d(E-h(T))\prod_td^2\x d^2\eta
\d(\dot{\x}_1-\o_1-j_{\x_1})
\d(\dot{\x}_2-\o_2-j_{\x_2})
\n \\\times
\d(\dot{\eta}_1-j_{\eta_1})
\d(\dot{\eta}_2).
\l{33}\ea

Note now that the $\x, \eta$ variables are contained in $r_c$ only:
$$
r_c= r_c (\x_1, \eta_1, \eta_2).
$$
This means that the action of the operator $\h{j}_{\x_2}$ gives
identical to zero contributions into perturbation theory series. And,
since $\h{e}_{\x_2}$ and $\h{j}_{\x_2}$ are conjugate operators, see
(\r{31}), we can put
$$
j_{\x_2}=e_{\x_2}=0.
$$
This conclusion ends the reduction:
\be
\h{K}(j,e)=\int^T_0 dt (\h{j}_{\x_1}\h{e}_{\x_1}+
\h{j}_{\eta_1}\h{e}_{\eta_1}),
\l{34}\ee
\be
e_c=e_{\eta_1}\f{\pa r_c}{\pa \x_1}-e_{\x_1}\f{\pa r_c}{\pa \eta_1}.
\l{34'}\ee
The measure has the form:
\be
DM(\x, \eta)=\d(E-h(T))d\x_2 d\eta_2\prod_td\x_1 d\eta_1
\d(\dot{\x}_1-\o_1-j_{\x_1})
\d(\dot{\eta}_1-j_{\eta_1})
\l{35}\ee
since $V=V(r_c, e_c, \x_1)$ is $\x_2$ independent and
$$
\int \prod_t dX(t)\d(\dot{X})=\int dX(0).
$$

\section{Perturbations}\0

One can see from (\r{35}) that the reduction can not solve the H-atom
problem completely: there are nontrivial corrections to the orbital
degrees of freedom $\x_1,\eta_1$. By this reason we should consider
the expansion over $\h{K}$.

Using last $\d$-functions in (\r{35}) we find, see also \C{yaph}
(normalizing $\R(E)$ on the integral over $\x_2$):
\be
\R(E)=\int^\infty_0 dT e^{-i\h{K}(j,e)}\int dM e^{-iV(r_c,e)},
\l{41}\ee
where
\be
dM=\f{d\x_1 d\eta_1}{\o_2(E)}.
\l{42}\ee
The operator $\h{K}(j,e)$ was defined in (\r{34}) and
\ba
V(r,e)=-s_0(r)+\int^T_0 dt[\f{1}{((r_c+e_c)^2+
r_c^2e_{\x_1}^2)^{1/2}}-
\n \\
-\f{1}{((r_c-e_c)^2+r_c^2e_{\x_1}^2)^{1/2}}+2\f{e_c}{r_c}]
\l{a}\ea
with $e_c,~e_{\x_1}$ defined in (\r{34'}, \r{32}) and
\be
r_c(t)=r_c(\eta_1 +\eta(t), \bar{\eta}_2(E,T), \x_1+\o_1(t)
+\x(t)),~~E\equiv h(\eta_1 +\eta(T), \bar{\eta}_2),
\l{43}\ee
where $\bar{\eta}_2(E,T)$ is the solution of equation $E=h$.

The integration range over $\x_1$ and $\eta_1$ is as follows:
\be
0\leq \x_1 \leq 2\pi,~~-\infty \leq \eta_1 \leq +\infty.
\l{45}\ee
First inequality defines the principal domain of the angular variable
$\vp$ and second ones take into account the clockwise and
anticlockwise motions of particle on the Kepler orbits.

We can write:
\be
\R(E)=\int^\infty_0 dT \int dM :e^{-iV(r_c,\h{e})}:
\l{47}\ee
since the operator $\h{K}$ is linear over $\h{e}_{\x_1},
\h{e}_{\eta_1}$.  The colons means `normal product' with operators
staying to the left of functions and $V(r_c,\h{e})$ is the functional
of operators:
\be
2i\h{e}_c=\h{j}_{\eta_1}\f{\pa r_c}{\pa \x_1}-
\h{j}_{\x_1}\f{\pa r_c}{\pa \eta_1},~~2i\h{e}_{\x_1}=\h{j}_{\x_1}.
\l{48}\ee
Expanding $V(r_c, \h{e})$ over $\h{e}_c$ and $\h{e}_{\eta_1}$ we
find:
\be
V(r_c,\h{e})=-s_0(r_c) +2\sum_{n+m \geq 1}C_{n,m}\int^T_0 dt
\f{\h{e}_c^{2n+1}\h{e}_{\eta_1}^m}{r_c^{2n+2}},
\l{46}\ee
where $C_{n,m}$ are the numerical coefficients. We see that the
interaction part presents expansion over $1/r_c$ and, therefore, the
expansion over $V$ generates an expansion over $1/r_c$.

In result,
\be
\R(E)=\int^\infty_0 dT \int dM \{e^{is_0 (r_c)} +
B_{\x_1}(\x_1, \eta_1) +
B_{\eta_1}(\x_1, \eta_1)\}.
\l{49}\ee
The first term is the pure quasiclassical contribution and last ones
are the quantum corrections. Using results of \C{yaph} functionals
$B$ are the total derivatives:
\be
B_{\x_1}(\x_1, \eta_1)=\f{\pa}{\pa \x_1}b_{\x_1}(\x_1, \eta_1),~~
B_{\eta_1}(\x_1, \eta_1)=\f{\pa}{\pa \eta_1}b_{\eta_1}(\x_1, \eta_1).
\l{410}\ee
This means that the mean value of quantum corrections in the $\x_1$
direction are equal to zero:
\be
\int^{2\pi}_0 d\x_1 \f{\pa}{\pa \x_1}b_{\x_1}(\x_1, \eta_1) =0
\l{411}\ee
since $r_c$ is the closed trajectory independently from initial
conditions, see (\r{214}).

In the $\eta_1$ direction the motion is classical:
\be
\int^{+\infty}_{-\infty} d\eta_1 \f{\pa}{\pa \eta_1}
b_{\eta_1}(\x_1, \eta_1)=0
\l{412}\ee
since (i) $b_{\eta_1}$ is the series over $1/r_c^2$ and (ii) $r_c
\rar \infty$ when $|\eta_1| \rar \infty$. Therefore,
\be
\R(E)=\int^\infty_0 dT \int dM e^{is_0 (r_c)}.
\l{413}\ee
This is the desired result.

Noting that
$$
s_0 (r_c)= kS_1 (E),~~k=\pm 1, \pm 2,...,
$$
where $S_1 (E)$ is the action over one classical period $T_1$:
$$
\frac{\partial S_1 (E)}{\partial E}=T_1 (E),
$$
and using the identity \C{manj}:
$$
\sum^{+\infty}_{-\infty} e^{inS_1 (E)} =
2\pi \sum^{+\infty}_{-\infty}\d (S_1 (E) - 2\pi n),
$$
we find:
\be
\R(E)=\pi \O \sum_{n} \d (E + 1/2n^2)
\l{416}\ee
where $\O$ is the zero-modes volume.

\section{Concluding remarks}\0

Our result (\r{413}) essentially uses the fact that the quantum
corrections are defined by the topology of the $G_\o$. Considering
$E\leq 0$ $G_\o$ has the topology of torus $G_\o=S^1\times S^1$ in
the $(l_0,h_0)$ plane \C{smale} and at $|l_0|=\infty$ this torus
degenerate to the circle with infinite radii. Therefore, because of
property (\r{412}) the mean value of quantum corrections lie on a
bifurcation line in the $(l_0,h_0)$ plane.

Absence of the angular corrections are due from the fact that the
classical trajectory in the Coulomb potential is closed independently
from $l$ and $h$.  This property reflects conservation of Runge-Lentz
vector, i.e. is the consequence of the hidden $O(4)$ symmetry
\C{pop}.

\vspace{0.2in}
{\Large \bf Acknowledgement}

I would  like to thank my colleagues in the Institute of Physics
(Tbilisi) for interesting discussions. The work was supported in part
by Georgian Academy of Sciences.

\end{document}